\documentclass[fleqn,usenatbib,linenumbers]{mnras}

\usepackage[T1]{fontenc}

\DeclareRobustCommand{\VAN}[3]{#2}
\let\VANthebibliography\thebibliography
\def\thebibliography{\DeclareRobustCommand{\VAN}[3]{##3}\VANthebibliography}

\usepackage{graphicx}	
\usepackage{amsmath}	
\usepackage{amssymb}	
\usepackage{graphicx}
\usepackage{graphicx}	
\usepackage{amsmath}	
\usepackage{amssymb}	
\usepackage{booktabs}
\usepackage{epsfig}
\usepackage{tablefootnote}
\usepackage{epstopdf}

\title[BL Lac spectral evolution and HBL component]{Broadband study of BL Lac during flare of 2020: Spectral evolution and emergence of HBL component}

\author[Prince et al.]{
Raj Prince\thanks{E-mail: raj@cft.edu.pl}
\\
Center for Theoretical Physics, Polish Academy of Sciences, Al.Lotnikov 32/46, PL-02-668 Warsaw, Poland\\
}

\date{Accepted XXX. Received YYY; in original form ZZZ}

\begin{document}
\label{firstpage}
\pagerange{\pageref{firstpage}--\pageref{lastpage}}
\maketitle

\begin{abstract}
   BL Lacertae (BL Lac) is categorized as TeV blazar and considered as a possible source of astrophysical neutrinos. In 2020, the brightest X-ray ﬂare ever detected from it. A detailed study can answer many puzzling questions related to multiband emissions and fast-flux variability often seen in this kind of source. We have performed the temporal and spectral analysis of the brightest ﬂare. The variability is characterized by the fractional variability amplitude and the variability time. We found that the source has crossed all its previous limits of ﬂux and reached to a maximum ever seen from it in optical and X-rays. 
   It is highly variable in X-rays with fractional variability above 100\% (1.8397$\pm$0.0181) and the fastest variability time of 11.28 hours within a day.
   The broadband light curves correlation with X-ray suggest a time lag of one day.
   A broadband SED modeling is pursued to understand the possible physical mechanisms responsible for broadband emission.
Modeling requires two emission regions located at two different sites to explain the low and high ﬂux states. A significant spectral change is observed in the optical-UV and X-ray spectrum during the high state, which eventually leads to shifts in the location of the synchrotron peak towards higher energy, suggesting an emergence of a new HBL component.
 \end{abstract} 

%

\section{Introduction}
Active galactic nuclei or AGN are the centers of an active galaxy. It consists of three components: a supermassive black hole (SMBH), an accretion disk surrounding the SMBH, and highly relativistic jets perpendicular to the accretion disk plane. The AGN are classified in various classes depending upon their jet orientation with respect to the observer. The classification was first proposed by \citet{Urry_1995} under the AGN unification scheme. In the AGN unification model, sources with relativistic jets highly oriented towards the observer within a few degrees are categorized as a blazar. Jets produce the Doppler boosted non-thermal highly collimated emission along the axial direction.
The high apparent luminosity due to Doppler boosting makes these sources highly luminous in the universe, and eventually, they can outshine their host galaxy(\citealt{Hartman_1999}, \citealt{Abdo_2010}). It is believed that the jet is powered by the SMBH present at the center of a galaxy (\citealt{Lynden-Bell_1969}). The mass of SMBH in AGN or blazar are generally falling between 10$^{6}$-10$^{10}$M$_{\odot}$. The observed spectral energy distribution (SED) of blazar show two broad peaks in low and high energy bands. The low energy peak covers the optical-UV and soft X-ray emission, and the high energy peak explains the X-ray and $\gamma$-ray emission from the source. With complete unanimity, the synchrotron process is accepted to explain the low energy peak of the SED, where the emission is produced by electrons gyrating along the magnetic field lines.
However, the high energy peak is a bit controversial because of the degeneracy in the models. The inverse-Compton (IC) scattering of low energy photons by the high energy electrons or the proton synchrotron or proton-photon interactions (\citealt{Bottcher_2013}) can produce the high energy peak of the SED. The model involves electrons are referred to as leptonic, and with protons are named as a hadronic model. In the leptonic scenario, the low energy seed photons for the IC scattering could be synchrotron photons produced inside the jet, known as synchrotron-self Compton (SSC; \citealt{Sikora_2009}), or from outside the jet ( broad-line region, dusty torus, accretion disk, etc.) identified as external Compton (EC; \citealt{Dermer_92}; \citealt{Sikora_94}).
In a proton-photon scenario, one would also expect to see high-energy neutrinos along with high-energy $\gamma$-ray.  
Blazars are further classified in sub-classes based on the presence or absence of emission lines in their optical spectra. Sources with strong emission lines are classified as flat-spectrum radio quasars (FSRQ) and with weak or no emission lines as BL Lacertae (\citealt{Stickel1991}, \citealt{Weymann1991}).
Further, the BL Lacertae is sub-divided based on the location of the synchrotron peak in their spectral energy distribution (SED). 
The synchroton peak observed at $\leq$10$^{14}$Hz is classified as low BL Lac (LBL), peak between 10$^{14}$ $\leq$ 10$^{15}$Hz as intermediate BL Lac (IBL), and $\geq$10$^{15}$Hz as high BL Lac (HBL) type (\citealt{Padovani1995}). Later, based on the broadband SED modeling \citet{Abdo2010} classified all the AGN into three categories namely, low synchrotron peaked blazars (LSP; $\nu_{peak}$ $\leq$10$^{14}$Hz), intermediate synchrotron peaked blazars (ISP; 10$^{14}$ $\leq$ $\nu_{peak}$ $\leq$ 10$^{15}$Hz), and high synchrotron peaked blazars (HSP; $\nu_{peak}$ $\geq$10$^{15}$) .
Blazar, in general, got more attention because of their variable nature and tendency to show stochastic flare across the wavebands. The variability observed in blazar has a wide spectrum in time ranging from minutes to several days (\citealt{Heidt_1996}, \citealt{Ulrich_1997}). The fast-flux variability time directly probes the inner part of the jet close to the SMBH, where the emissions are produced, which are difficult to probe otherwise by any currently available imaging telescopes. The variability study is important to understand the physics happening close to the SMBH and eventually jet launching and the particle acceleration mechanisms.
The fast-flux variability is not limited to a particular band but rather observed across the wavebands, and the recent development in the simultaneous data collection over different wavebands reveals the involvement of highly complex nature and multi-scale physics in blazar. 

BL Lacertae (BL Lac) is located at redshift, z = 0.069 (\citealt{Miller1978}). Based on synchrotron peak location, it is classified as intermediate BL Lac (IBL; \citealt{Ackermann2011}) and later as low BL Lac (LBL; \citealt{Nilsson2018}) type blazar. The recent classification presented by \citet{Hervet2016} based on the kinematic features of radio jets confirms the BL Lac as an IBL type blazar. BL Lac is known for its broadband variability and various multi-wavelength campaign have been carried out by several authors (\citealt{Hagen-Thorn2002}, \citealt{Marscher2008}, \citealt{Raiteri2009,Raiteri2013}, \citealt{Wehrle2016}, \citealt{2019A&A...623A.175M}), to understand the nature of this source.

Many studies in the past suggest that despite the complex nature and high flux variability, the SED of blazars shows a stable synchrotron and IC peak. However, a few sources are exceptional, where it is noted that the synchrotron peak can shift towards higher energy during the high flux state
(\citealt{Morini1986}, \citealt{Giommi1990}, \citealt{Raiteri15}, \citealt{kush18}, \citealt{Kapanadze_2018}). In this article, we addressed such findings also observed in BL Lac during the 2020 flaring state. The order of presentation is following, in section 2, we discussed the observation and data reduction from the broadband telescopes, section 3 is dedicated to the detailed X-ray study, and section 4 for $\gamma$-ray study, in section 5, we present the broadband light curves, and section 6 describes the spectral change seen in optical-UV and X-ray during the flaring state. In section 7, we discussed the temporal variability of the source, and in section 8, the broadband SED modeling, section 9 used to describe the shift observed in synchrotron peak position, and finally, the summary of the work.
 
\section{Multiwavelength Observations and Data Analysis}
 BL Lac showed strong activity in X-ray observed by Swift-XRT, and simultaneously it was also reported to have elevated $\gamma$-ray flux detected by Fermi-LAT telescope. The source was simultaneously monitored in ultraviolet and optical wavebands also by Swift-UVOT. 
To understand the nature of emission processes and high flux variability, multiband coverage is required. The simultaneous coverage of $\gamma$-ray from {\it Fermi-LAT}, X-ray from {\it Swift-XRT} and ultraviolet and optical from {\it Swift-UVOT} will provide a great opportunity to study the temporal and spectral properties of the flare and, in general, the source. 

\subsection{\textit{Fermi}-LAT}
LAT (Large Area Telescope) is a $\gamma$-ray instrument onboard the Fermi satellite launched by NASA in 2008. Together it is known as Fermi $\gamma$-ray space telescope. The sole purpose of the instrument is to explore the $\gamma$-ray universe along with the AGILE and DAMPE.  \textit{Fermi}-LAT is more sensitive in the energy range between 100 MeV $-$ 300 GeV.
Blazars highly populate the $\gamma$-ray sky, and the recently published 4FGL catalog (\citealt{Abdollahi_2020}) revealed more than five thousand sources in $\gamma$-ray.
Fermi has a large field of view (FoV) of an order of 2.4 sr (\citealt{Atwood_2009}). It is an all-sky monitoring instrument with a scanning period of $\sim$96 minutes in both northern and southern sky, and in total, it takes $\sim$3 hours to scan the whole sky. 

BL Lac is continuously monitored by the Fermi-LAT since its beginning and appeared in Fermi third (\citealt{Acero_2015}) and fourth source catalogs (\citealt{Abdollahi_2020}). 
To characterize its $\gamma$-ray behavior during the flaring state in 2020, three months of data were collected and studied. 
The analysis is done following the standard procedure available at Fermi science tools\footnote{https://fermi.gsfc.nasa.gov/ssc/data/analysis/documentation/}. The further detailed analysis can be found in \citet{Prince_2018}. One-day and 12 hours binned light curves are produced for better understanding of flare's temporal behavior, with default Logparbola spectral model as published in 4FGL catalog, and shown in Figure \ref{fig:gam-ray}. 
To have better significance,
the data points shown in the light curves are chosen with TS > 10, which corresponds to approximately 3$\sigma$ (\citealt{Mattox_1996}) confidence level. 

\subsection{Neil Gehrels Swift Observatory}
Swift is a space-based observatory mainly focused on detecting the transient phenomenon in the universe with on-board three instruments, namely, X-ray telescope (XRT), ultraviolet optical telescope (UVOT), and the burst alert telescope (BAT). 
The blazar BL Lac is observed in all the instruments simultaneously for an extended period along with the Fermi $\gamma$-ray space telescope.

\subsubsection{Swift-XRT} 
The XRT has an energy range of 0.3$-$10.0 keV and
continuously monitoring the source BL Lac since the beginning of its operation. We produced a long X-ray light curve and found that the flare seen during October 2020 is the brightest flare ever detected from this source.
To reduce the X-ray data, we followed the standard procedure as used by the community. The \texttt{XRTPIPELINE} with the latest version of calibration file (CALDB) produced the cleaned event files, which are further used to create a source and background region using task \texttt{XSELECT}. 
A circular region of 10 and 30 arcsec is chosen around the source and away from the source for the source and background region, respectively.
Proper redistribution matrix files (RMF) and the ancillary response files (ARF) are used for further analysis. RMF, ARF, and background files are combined with the source spectrum through the task \texttt{grppha} where the spectra are grouped, with 30 counts per bin,
to have sufficient counts in each bin. 
The grouped spectra is then added to \texttt{XSPEC} (\citealt{1996ASPC..101...17A}) for the spectral analysis. Various spectral models have been used to model the spectra with a fixed galactic absorption column density, $n_H$ = 0.18 $\times$ 10$^{22}$ cm$^{-2}$(\citealt{Kalberla_2005}).
More than one observation available in a particular period are combined in 
 \texttt{addspec}\footnote{https://heasarc.gsfc.nasa.gov/ftools/caldb/help/addspec.txt}, where RMF and ARF files for different observations are also added. Similarly, the background spectra
from different observations are added through \texttt{mathpha}\footnote{https://heasarc.gsfc.nasa.gov/ftools/caldb/help/mathpha.txt}. The combined X-ray spectrum for flaring and low state is produced for further comparison and finally for the broadband SED modeling.

\subsubsection{Swift-UVOT}
Simultaneous observations in optical-UV with X-ray and $\gamma$-ray provides an unprecedented opportunity to understand the various physical mechanisms happening in the interior of the source. 
Swift-UVOT (\citealt{Roming_2005}) observe the source in quite a wide range of wavelength with
three optical (U, B, V) and three UV (W1, M2, W2) filters.
The source and background regions of 5 and 15 arcsec are chosen from the image files around the source and away from the source, respectively, and task \texttt{UVOTSOURCE} is used to extract the magnitude and flux values. In order to correct the magnitudes and fluxes from the galactic extinction, the reddening E(B-V) = 0.28 from \citet{Schlafly_2011} and the extinction ratios (A$_{\lambda}$/E(B-V)) from the \cite{Giommi_2006} are used. The corrected magnitudes are converted to flux by using the zero points from the \citet{Breeveld_2011} and the conversion factors from the \citet{Poole_2008}.

\section{Results}
\subsection{X-ray study}
\begin{figure*}
    \centering
    \includegraphics[scale=0.5]{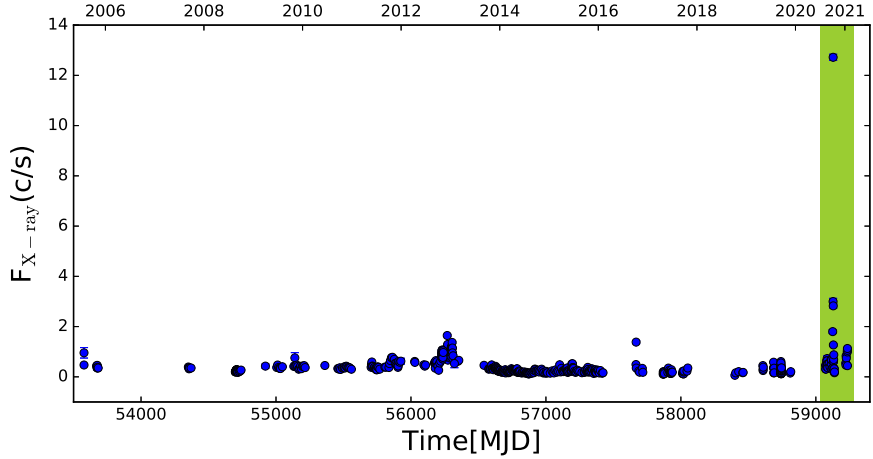}
    \caption{Historical 0.3-10.0 keV X-ray light curve from Swift-XRT. The selected shaded region is chosen for the further/detailed study. }
    \label{fig:total-xray}
\end{figure*}

We produced the historical X-ray light curve of BL Lac, using Swift-XRT observations in the energy range 0.3-10.0 keV as shown in Figure \ref{fig:total-xray}. The source is observed to be in the brightest state in October 2020, with a count rate of 13 counts/sec, six times higher than the historical average counts rate.
The active state is marked with a green color patch
in Figure \ref{fig:total-xray}, and the estimated flux and spectral index are shown in Figure \ref{fig:xray-flare}.
 Based on the flux values, the entire period in Figure \ref{fig:xray-flare} is divided into two parts; low and high state. 
 The high state is defined between 2020 October 1 to 2020 October 12 with the highest flux 3.44$\times$10$^{-10}$ erg cm$^{-2}$ s$^{-1}$ in energy 0.3-10.0 keV, ever seen from the source.
 The spectral variation during the active state is plotted in Figure \ref{fig:xray-index}, suggesting a clear "softer-when-brighter" trend. 

\subsection{X-ray spectral analysis}
\begin{figure*}
    \centering
    \includegraphics[scale=0.53]{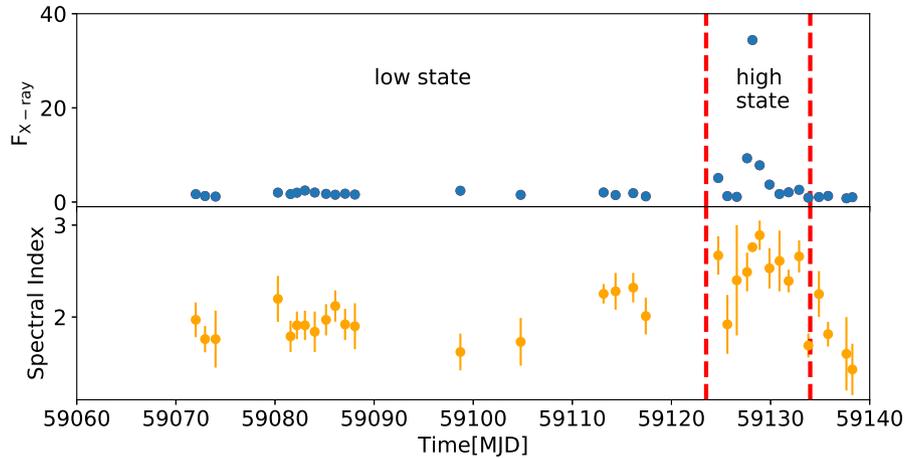}
    \caption{This shows the zoomed version of the shaded region from Figure \ref{fig:total-xray}. The entire period is divided in low and high state separated by the vertical dashed red line. The X-ray flux points are derived for 0.3-10.0 keV in units of 10$^{-11}$ erg cm$^{-2}$ s$^{-1}$.}
    \label{fig:xray-flare}
\end{figure*}

\begin{figure}
    \centering
    \includegraphics[scale=0.45]{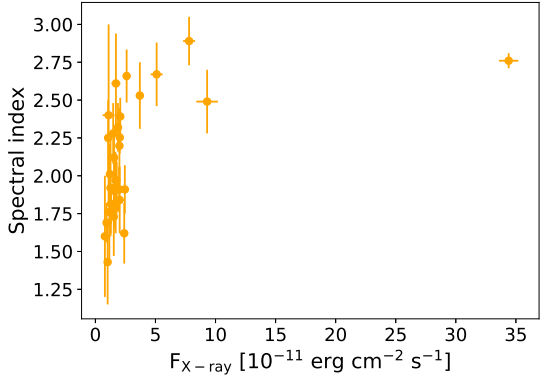}
    \caption{X-ray spectral index with respect to X-ray flux from 0.3-10.0 keV are shown here. A softer-when-brighter trend with respect to flux is observed. }
    \label{fig:xray-index}
\end{figure}

\begin{figure}
    \centering
    \includegraphics[scale=0.3,angle=-90]{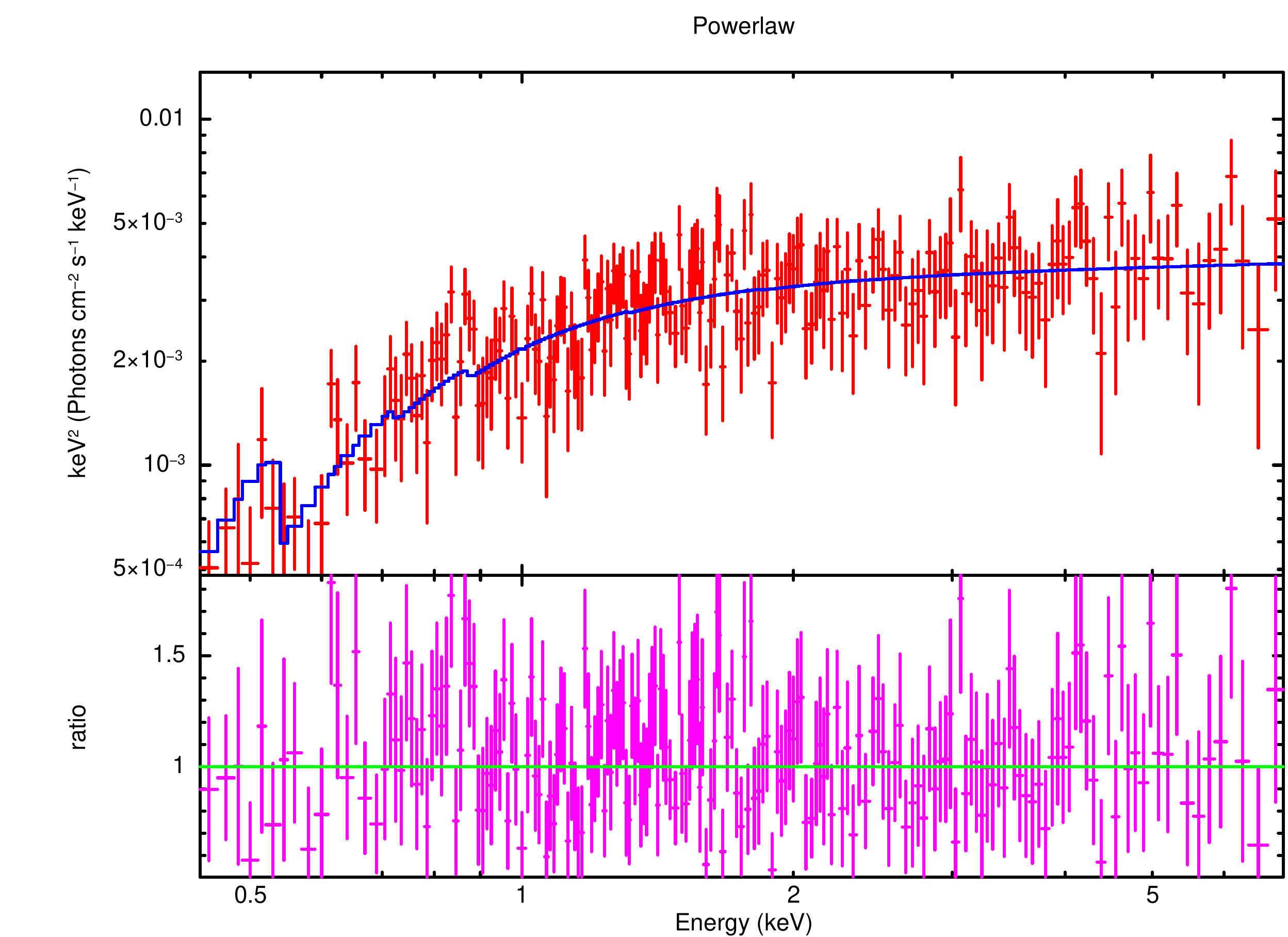}
    \includegraphics[scale=0.3,angle=-90]{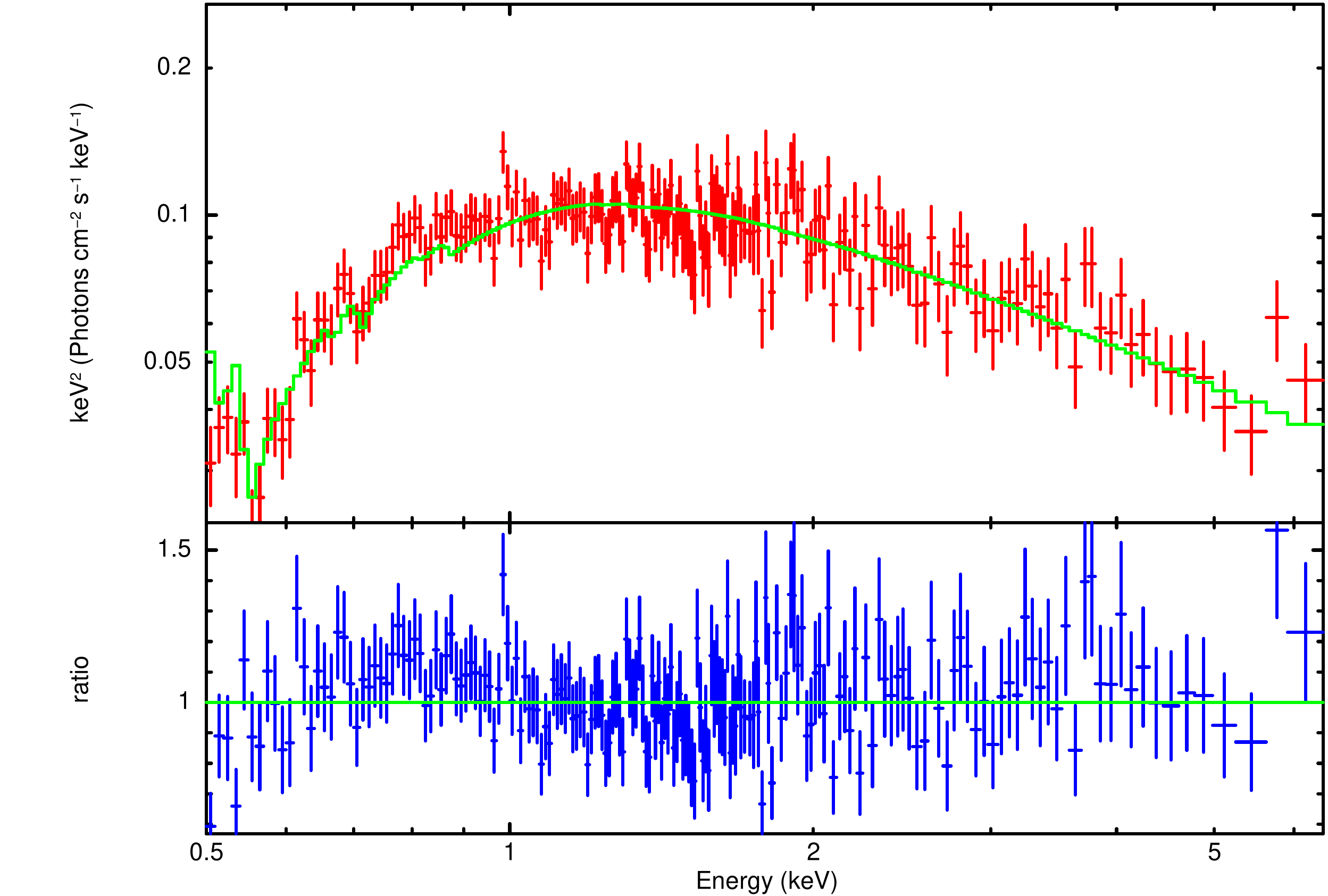}
    \caption{X-ray spectrum generated for total period of low state and an individual observation 00034748023 from the high state. Upper panel: total low state spectrum, Lower panel: individual observation 48023. The fitted parameters are presented in Table \ref{tab:tabspec}. }
    \label{fig:xray-sed}
\end{figure}

To examine the spectral evolution, the X-ray spectrum of both low and high states is produced and compared.
The flux level is constant over time in a low state (see Figure \ref{fig:xray-flare}), and hence a common spectrum is produced after combining the spectra from individual observations. 
The high state is comprised of few individual observations, and their spectra are produced and compared among them and with the low state. 
The spectra are fitted with two different spectral models namely, power-law (PL) and log-parabola (LP). The functional form of those are following,
\begin{equation}
   F(E) = K E^{-\Gamma} 
\end{equation}
where F(E) is flux at energy E, K is the normalization, and the $\Gamma$ is spectral index.
\begin{equation}
   F(E) = K (E/E_{ref})^{-\alpha + \beta log(E/E_{ref})} 
\end{equation}
where E$_{ref}$ is called the reference energy which is fixed at 1.0 keV, $\alpha$ is the index, and $\beta$ is curvature index of LP spectrum.
Our analysis showed that for every individual observation in low state as well as in high state, PL is the best fit model. The parameters corresponding to the spectrum analysis is provided in Table \ref{tab:tabspec}. 
Further, we used the F-test statistics between the PL and LP models, and the results suggest that distinction can not be made by the F-test statistics since the models fit showed higher null hypothesis probability. 
Furthermore, it is noted that all the spectra correspond to 18 Swift observations during low-state (see Figure \ref{fig:xray-flare}), showed a harder spectrum compared to the individual observations of the high state, the fitted spectra are shown in Figure \ref{fig:xray-sed}.
The average spectral index obtained by combining all the observations in the low state is 1.94$\pm$0.06, which is harder compared to a single observation at the highest flux with spectrum 2.95$\pm$0.06, suggesting a "softer-when-brighter" trend as also seen from the individual observations in Figure \ref{fig:xray-flare} and \ref{fig:xray-index}. 
 In addition, the "softer-when-brighter" trend is seen the first time in BL Lac history of X-ray observations.
 The study by \citet{Wehrle2016} showed an opposite trend. The significant spectral change seen in Figure \ref{fig:xray-sed} is observed the first time in this source. Though the change in the NIR-optical-UV and X-ray spectra during 2007$-$2008 flares are studied in detail by \citet{Raiteri2009}.

\begin{table*}
    \centering
    \begin{tabular}{c|cc|ccc|cc}
    \hline
    \noalign{\smallskip}
 {\bf States} & PL & &LP& & &F-test & P-value \\
  & $\Gamma$& ${\chi}^2$ (dof)& $\alpha$& $\beta$ &${\chi}^2$(dof)  & &  \\
  \noalign{\smallskip}
  \hline \noalign{\smallskip}
 low state & 1.94$\pm$0.06 & 190.16(195)&1.97$\pm$0.11 & 0.05$\pm$0.18 & 189.92(194) & 0.24 & 0.62 \\
 \noalign{\smallskip} \hline \noalign{\smallskip}
 HS: 00034748023 & 2.95$\pm$0.06 & 370.20(321)& 3.06$\pm$0.29 & 0.13$\pm$0.34 & 369.81(320)& 0.34 & 0.56 \\
\noalign{\smallskip} 
  \hline 
    \end{tabular}
    \caption{ Modeled parameters for the X-ray spectrum during low state and of high state(HS). In low state, we have combined the spectrum from different observation however, in high state we have shown the spectrum corresponds to the brightest flux state.}
    \label{tab:tabspec}
\end{table*}

\subsection{$\gamma$-ray flare}
\begin{figure*}
\centering
\includegraphics[scale=0.6]{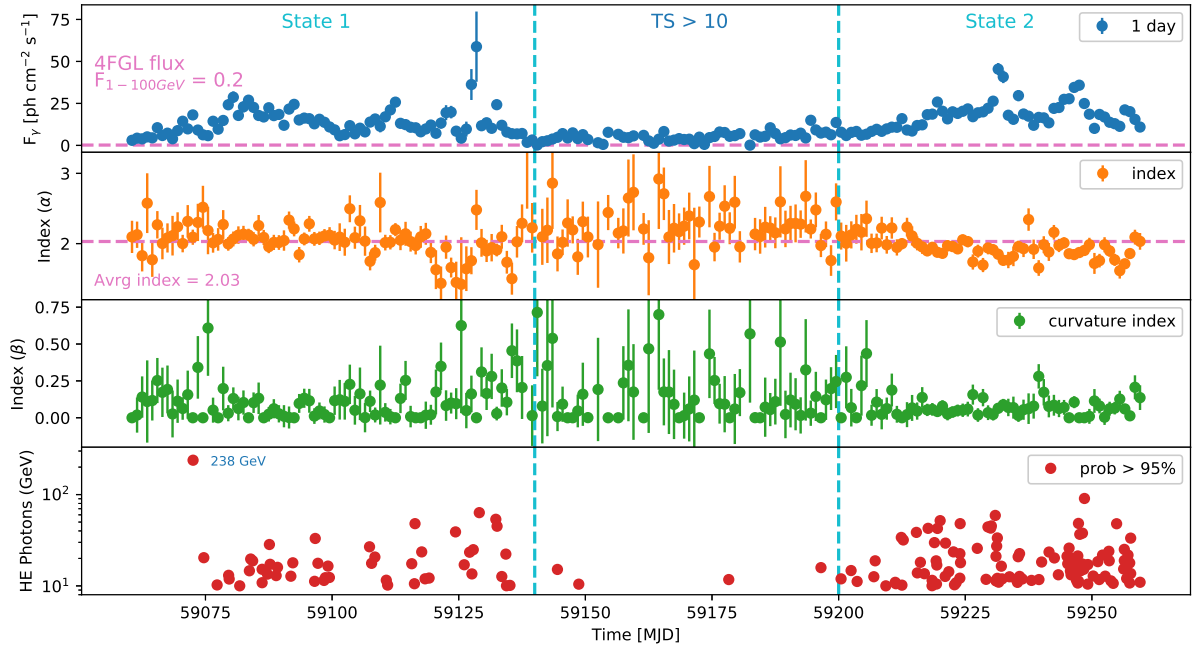}
\caption{The $\gamma$-ray behavior of BL Lac during the 2020 X-ray flaring state defined in Figure \ref{fig:total-xray}. The $\gamma$-ray light curve is also generated for longer period and a flaring episode is observed between MJD 59200 to 59300 and during this period X-ray is in low state. The $\gamma$-ray flux is in units of 10$^{-7}$ ph cm$^{-2}$ s$^{-1}$.}
\label{fig:gam-ray}
\end{figure*}

We analyzed the $\gamma$-ray data for the entire X-ray flaring period. In addition, more $\gamma$-ray data are collected to explore the flaring activity in $\gamma$-ray between MJD 59060--59260 (August 1, 2020, to February 15, 2021). 
One-day binned $\gamma$-ray light curves are generated as shown in the top panel of Figure \ref{fig:gam-ray}. Based on the average flux value, the total period is divided into two parts such as \texttt{state 1} and \texttt{state 2}, and the period is marked with a vertical dashed line. 
The flux level of \texttt{state 1} and \texttt{state 2} are much above the average 4FGL flux level, denoted by a horizontal dashed line.
The data points shown in the plot are above 3$\sigma$(TS$>$10) as generally considered for the $\gamma$-ray analysis.

Along with flux points, the spectral indices are extracted by using the default
log parabola spectral models are also shown in panels 2 and 3. 
The average spectral index is estimated as 2.03,  shown by a horizontal dashed line in panel 2. It is noted that during \texttt{state 1} most of the spectral indices are above the average value suggesting a softer behavior with increasing flux (softer-when-brighter). On the other hand, during \texttt{state 2} most of the indices are below the average spectral index hinting a harder behavior with increasing flux (harder-when-brighter). This behavior is later discussed in detail with flux versus index plot.

The highest energy $\gamma$-ray photons with respect to arrival times are plotted in panel 4 of Figure \ref{fig:gam-ray}. All photons have energy above 10 GeV and probability for being from the source above 95$\%$.
The low state-defined between MJD 59140--59200 have only a few high energy photons, whereas most of the high energy photons are observed during flaring \texttt{state 1} $\&$ \texttt{state 2}, suggesting a connection between the production of high energy photons and high flux state of the source.
During \texttt{state 1}, the highest energy photon detected is $\sim$ 238 GeV, however, during \texttt{state 2} few more than 100 GeV photons are observed.

\begin{figure*}
    \centering
    \includegraphics[scale=0.5]{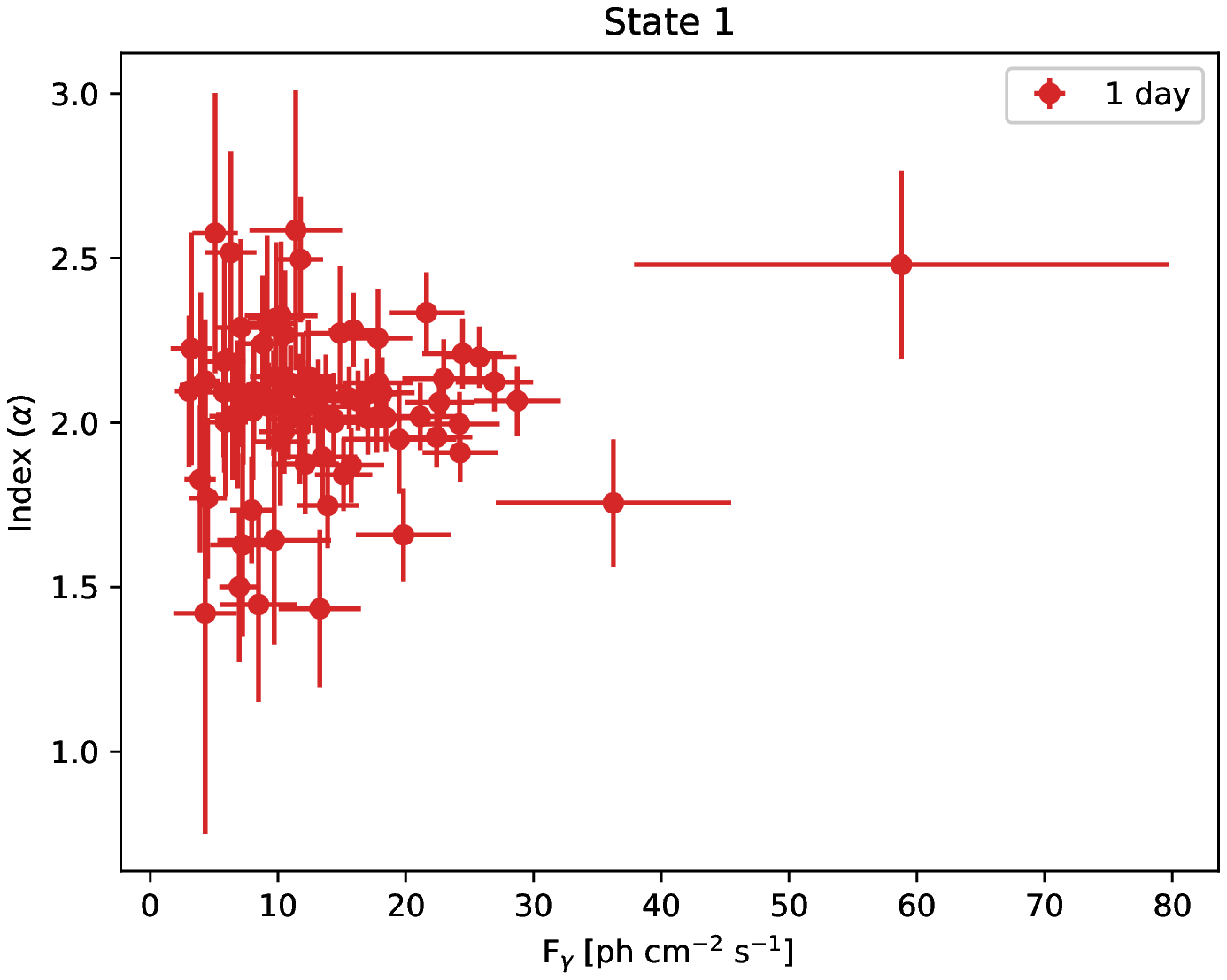}
    \includegraphics[scale=0.5]{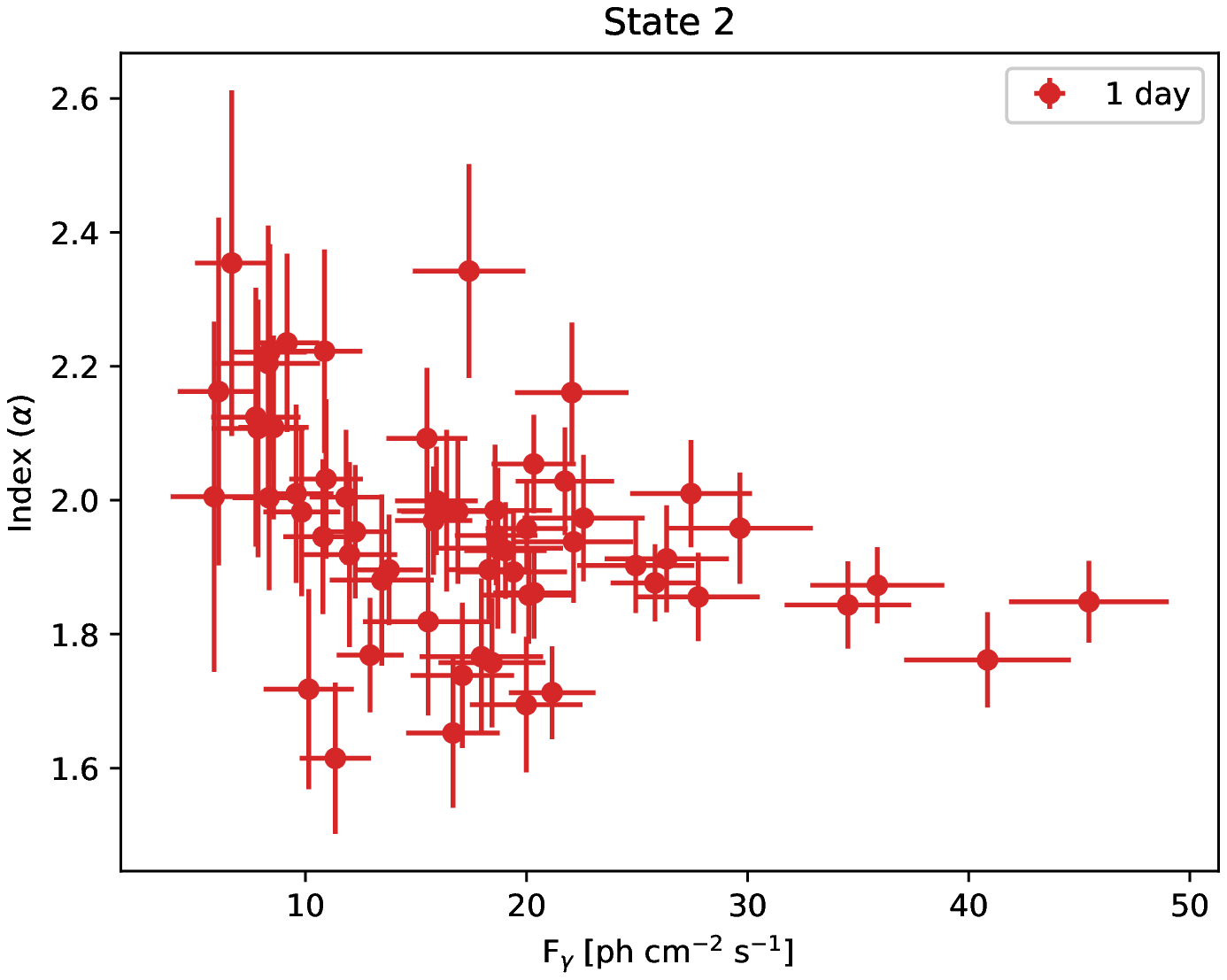}
    \caption{Plot show the flux versus index ($\alpha$) behavior of the source during \texttt{state 1} and \texttt{state 2}. }
    \label{fig:state-index}
\end{figure*}

The spectral behavior of \texttt{state 1 \& state 2} with respect to corresponding fluxes are shown in Figure \ref{fig:state-index}. 
The $\gamma$-ray spectral index during \texttt{state 1} (mostly X-ray flaring period) showed a mild trend of "softer-when-brighter", however, a "harder-when-brighter" trend is observed during \texttt{state 2}, a very common behavior of blazar.

Since the bright X-ray flare has a common period with the \texttt{state 1} of the $\gamma$-ray light curve, therefore, further study is focused only on \texttt{state 1}.

\section{Multi-wavelength Light Curves}
\begin{figure*}
    \centering
    \includegraphics[scale=0.5]{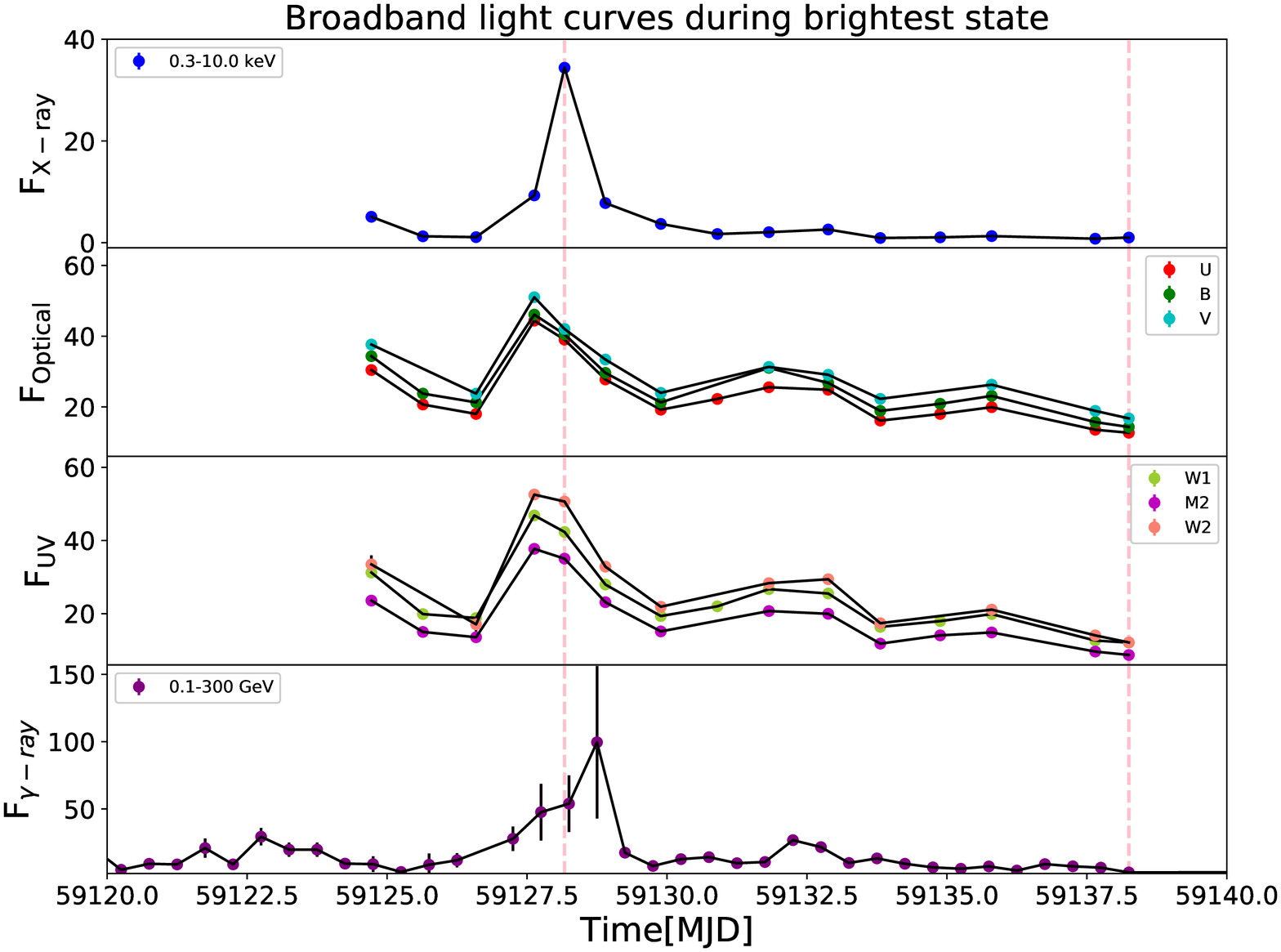}
    \caption{Broadband light curves of the flaring state observed in X-ray. X-ray, Optical, and UV fluxes are in units of 10$^{-11}$ erg cm$^{-2}$ s$^{-1}$. The $\gamma$-ray flux is in 10$^{-7}$ ph cm$^{-2}$ s$^{-1}$ unit. }
    \label{fig:MWL}
\end{figure*}

The broadband light curves of the X-ray flaring period along with optical-UV and $\gamma$-rays are shown in Figure \ref{fig:MWL}. The observed X-ray peak did not coincide with any of the waveband's emissions. 
BL Lac first peaked in optical-UV, then in X-rays followed by $\gamma$-ray, suggesting a small time lag amongst them.
During the post flare period, the light curves followed a similar trend and eventually settling-up in a low flux state.

\begin{figure*}
    \centering
    \includegraphics[scale=0.4]{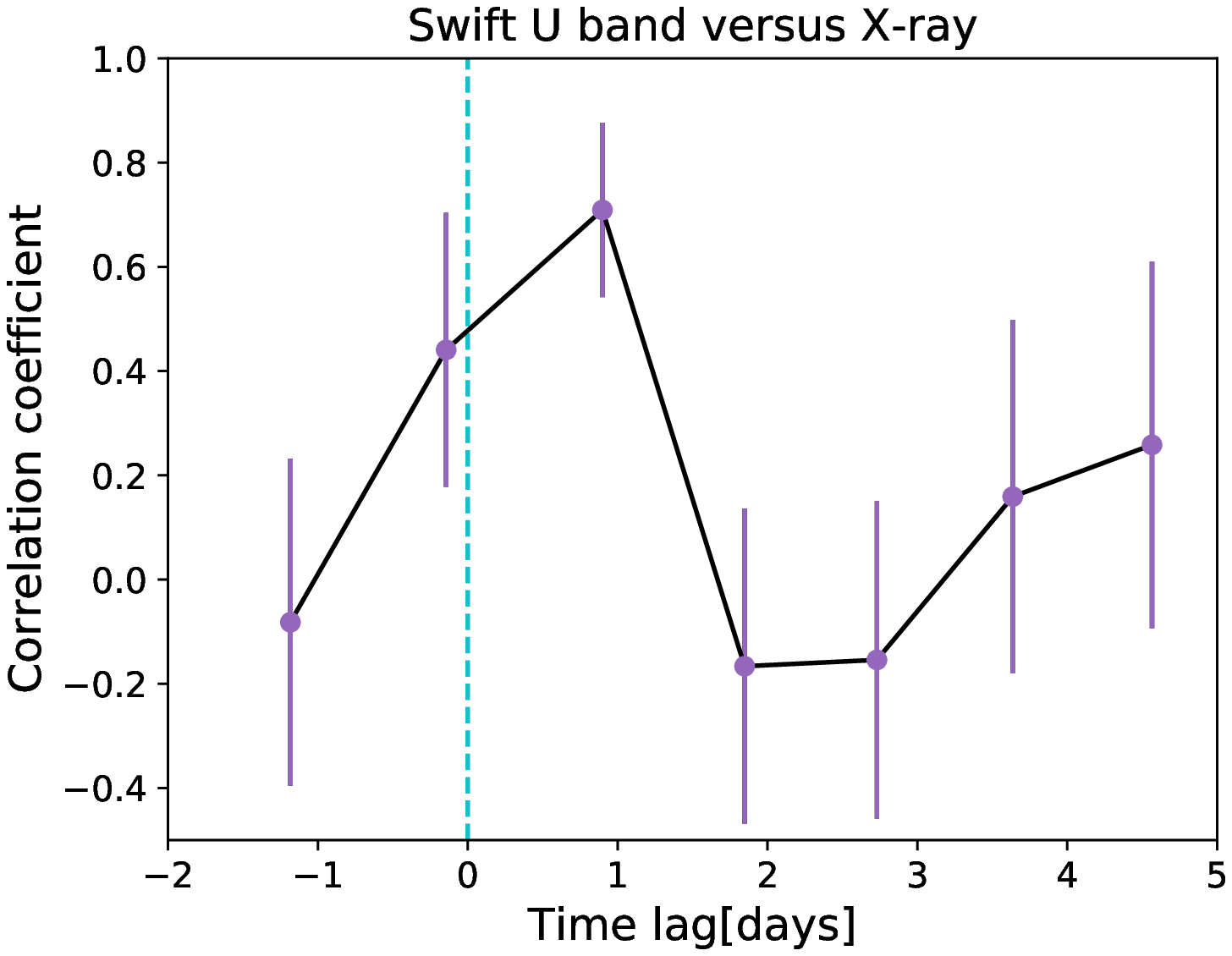}
    \includegraphics[scale=0.4]{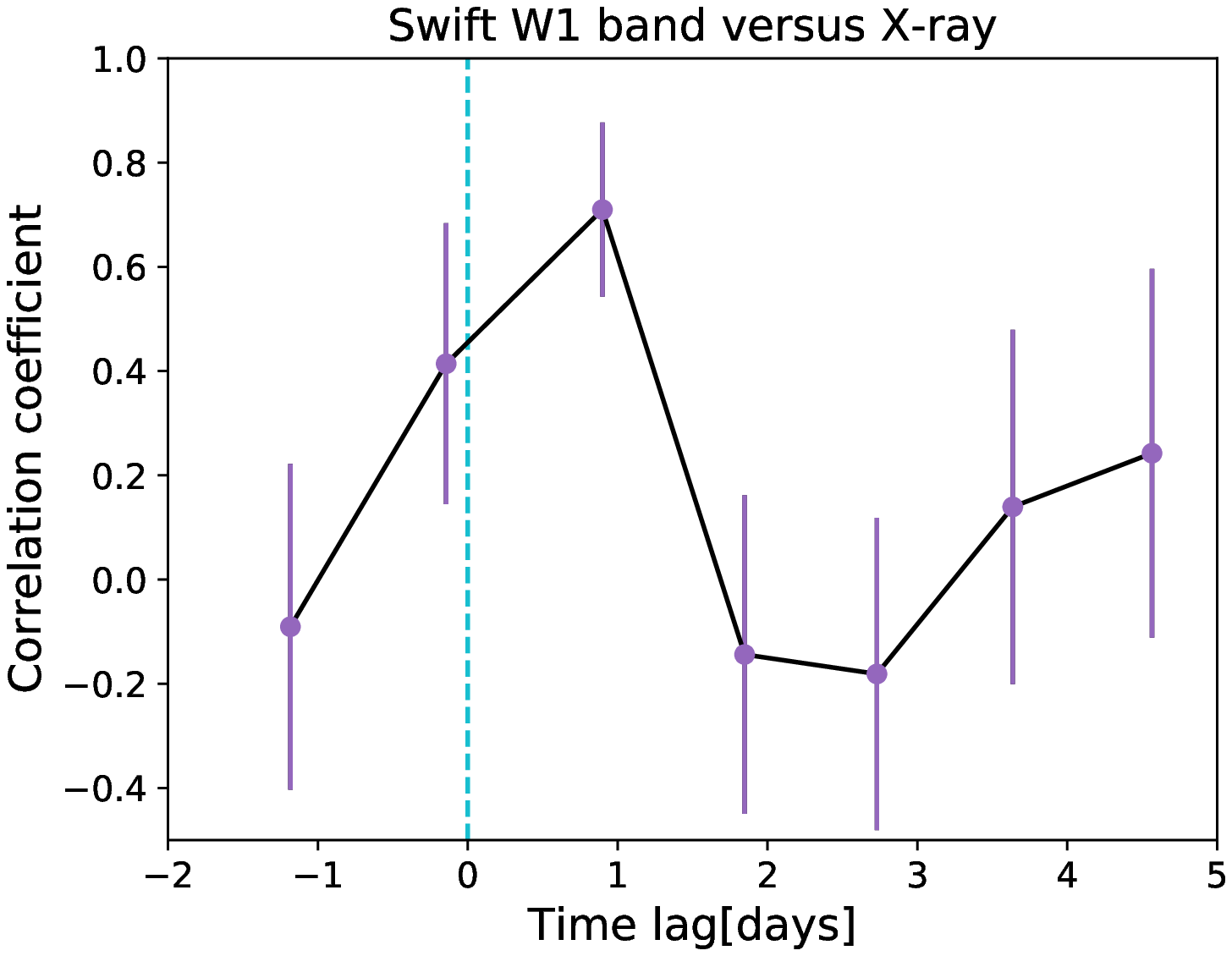}
    \includegraphics[scale=0.4]{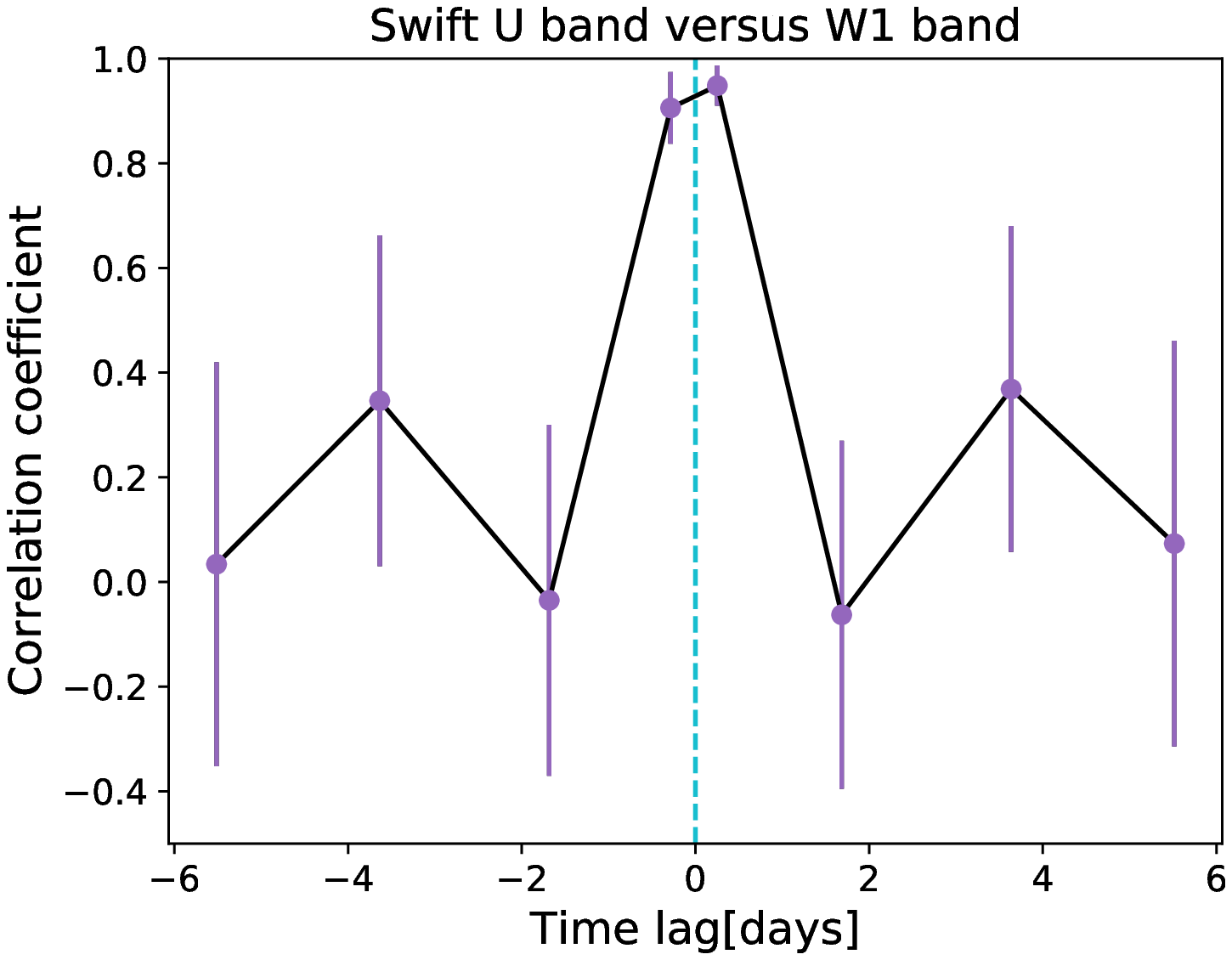}
    \includegraphics[scale=0.4]{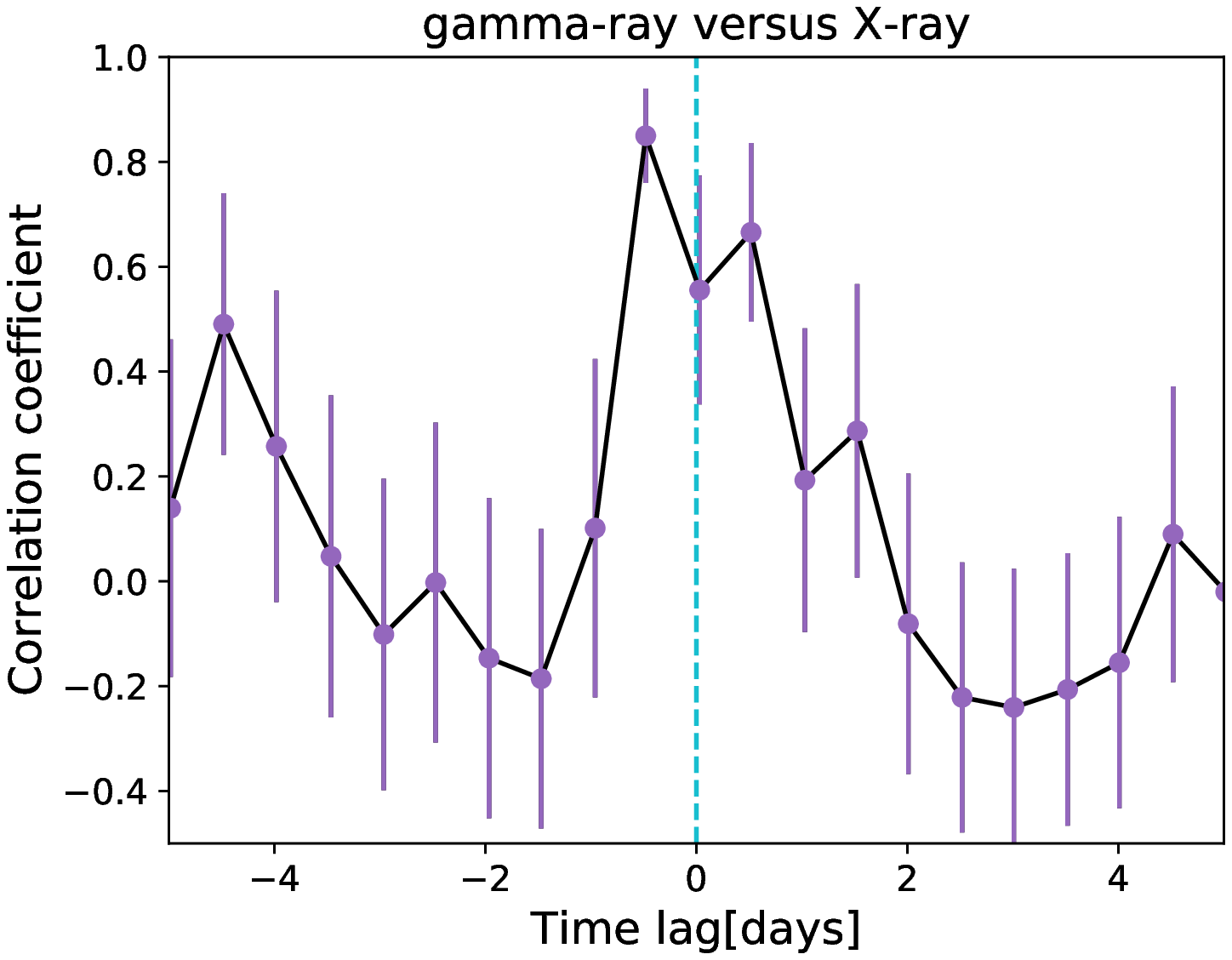}
    \caption{The correlation among various waveband.}
    \label{fig:dcf}
\end{figure*}

To examine the day's scale time lag evident in X-ray, optical-UV, and $\gamma$-ray, a broadband correlation study is performed using the discrete correlation (DCF) method \citep{Edelson_1988}. 
The details about the DCF can be found in \citep{Prince_2020}, and the results are shown in Figure \ref{fig:dcf}. 
The obtained results from the correlation study are in accordance with what was expected by eye inspection. 
The UV and optical emission lead the X-ray emission by one day. Similarly, X-ray leads the $\gamma$-ray emission by the same amount.
The optical-UV emission showed zero time lag as expected.  

\section{Spectral change in X-ray and optical-UV}
\begin{figure}
    \centering
    \includegraphics[scale=0.45]{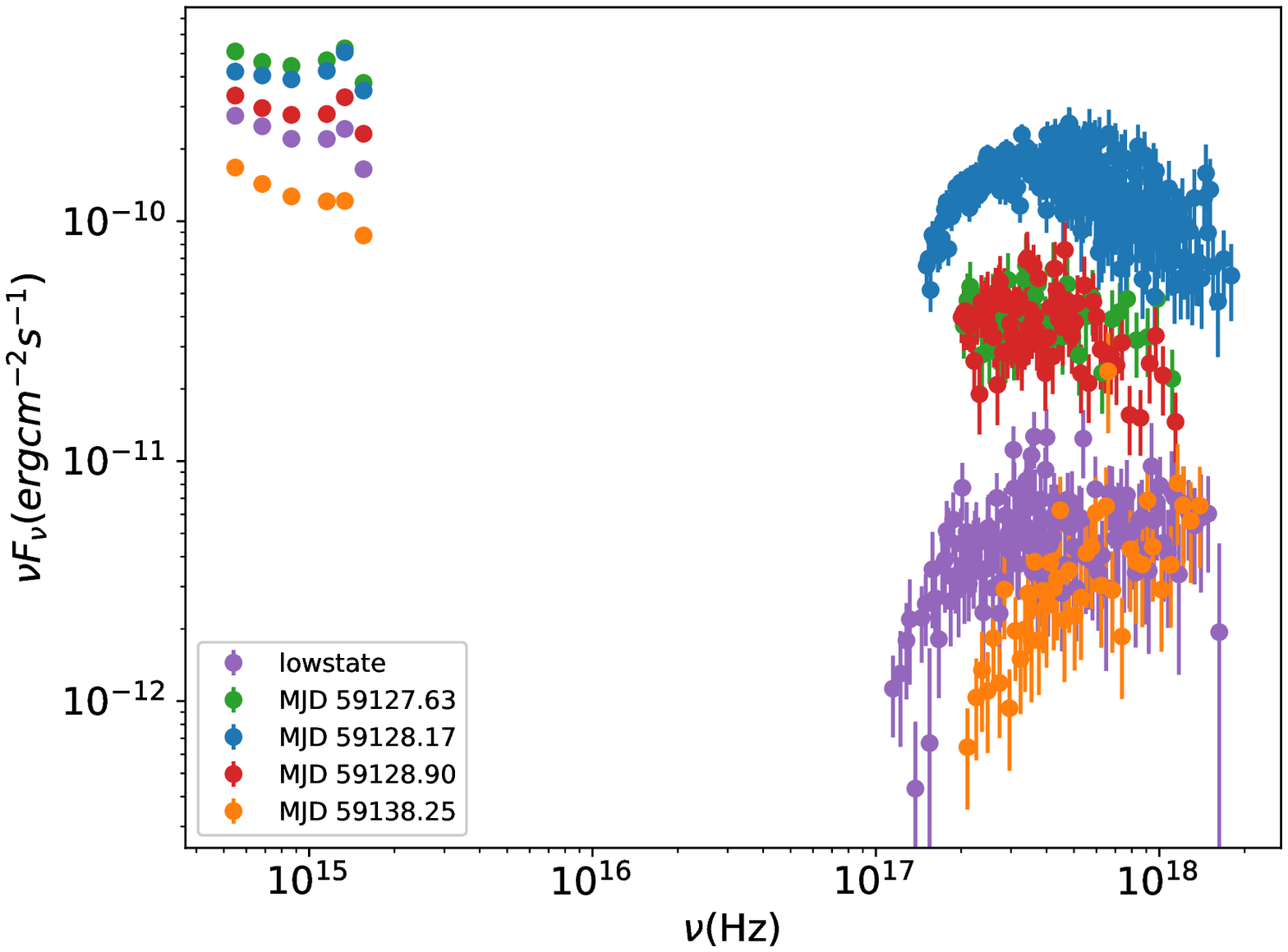}
    \caption{UVOT and X-ray spectral shape observed during low and high state. }
    \label{fig:xray-uv}
\end{figure}

To examine the X-ray and optical-UV spectral evolution between the low and high flux state, spectrums at various moments are derived. For the comparison purpose, spectrum at peak flux of the X-ray flare and other two sides of the peak flux is produced (Figure \ref{fig:MWL}).
X-ray peaked at MJD 59128.17, and the two side observations are taken from MJD 59127.63 and 59128.90. 
All the three observations have been considered as a part of a high flux state because their fluxes are above $\sim$7.5$\times$10$^{-11}$ erg cm$^{-2}$ s$^{-1}$. However, the peak flux is more than $\sim$4 times the fluxes of side observations. The spectrum is fairly steep for the two sides with spectral index 2.30$\pm$0.12 and 2.50$\pm$0.08, and significantly steep at peak with spectral index 2.95$\pm$0.01, suggesting a significant change in the X-ray spectrum within a duration of a day.  
Similarly, we also produced the spectrum for a significantly low flux state as marked by the vertical pink dashed line in Figure \ref{fig:MWL} at MJD 59138.25. The spectral index is found to be very hard compared to the X-ray peak with a value 1.37$\pm$0.19.
Additionally, we also analyzed the spectrum of the period just before the bright X-ray peak, where the source was in a low state with a very low count rate. We summed four consecutive observations taken at MJD 59113.15, 59114.36, 59116.14, and 59117.41 and produced a combined X-ray spectrum. The spectral index of the combined spectrum is found to be 2.13$\pm$0.10. Furthermore, the simultaneous UV/optical spectrum produced using UVOT showed a changing behavior from fairly steep in a low state to fairly flat in a high flux state and again significantly steep in a low state.
The combined effect of spectral changes seen in optical-UV and X-ray during low and high flux states (see Figure \ref{fig:xray-uv}) suggests a high energy shift in synchrotron peak, and more discussions are provided in section 8.
All the spectra are denoted by their corresponding observational times in Figure \ref{fig:xray-uv}. 

\section{Multiband Temporal Studies}

BL Lac is monitored across the entire electromagnetic spectrum by various grounds as well as space-based telescopes. The recent detection of the brightest X-ray flare in Swift-XRT and simultaneous coverage in UVOT and Fermi $\gamma$-ray space telescope provides a great opportunity to study the multiband temporal behavior.

Blazars are known for their significant stochastic variability on diverse time scales, which can be quantified by the excess variance ($\sigma_{XS}$), and the fractional RMS variability amplitude, F$_{var}$ (\citealt{Edelson_2002}). The excess variance, $\sigma_{XS}$, corresponds to intrinsic variability in blazar and is estimated by correcting the total observed light curve with measurement errors. Mathematically, F$_{var}$ is the square root of the $\sigma_{XS}$ normalized by the mean flux value, and its functional form and error are taken from \citet{Vaughan_2003}.

\begin{equation}
 F_{\rm var} = \sqrt{\frac{S^2 - err^2}{F^2}},
\end{equation}
where, $F$ denotes the mean flux value, S$^2$ is the total variance of the observed light curve, and err$^2$ is the mean square error in the observed flux. The functional form of the error on F$_{var}$ is defined in \citet{Prince_2019}. The fractional variability amplitude estimated for the various wavebands is depicted in Table \ref{tab:Fvar}. BL Lac is found to be the most variable in X-ray and $\gamma$-ray followed by UV and optical bands. 

\begin{table}
    \centering
    \begin{tabular}{ccc}
    Bands   & F$_{var}$ & t$_{var}$ (days)  \\
    \hline
    X-ray   & 1.84$\pm$0.02 & 0.47 \\
    U   &  0.38$\pm$0.01 & 1.23 \\
    B   &  0.35$\pm$0.01 & 1.41 \\
    V   &  0.56$\pm$0.01 & 1.42 \\
    W1   & 0.41$\pm$0.01 & 1.22 \\
    M2   &  0.67$\pm$0.01 & 1.02 \\
    W2  &  0.46$\pm$0.01  & 1.10 \\
    $\gamma$-ray & 0.91$\pm$0.13 & 0.36 \\
    \hline
    \end{tabular}
    \caption{Fractional variability and the fastest variability time in all the wavebands during X-ray flaring state.}
    \label{tab:Fvar}
\end{table}

Along with the strength of the variability, the variability time scale can also be estimated, using the expression from \citep{Zhang_1999}. 
The variability time is measured as a flux doubling time, where the flux changed from two or more than two factors in consecutive time intervals. The functional form of the expression is following
\begin{equation}
\centering
 t_{d} = \bigg|\frac{(F_1 + F_2)(T_2 - T_1)}{2(F_2 - F_1)}\bigg|
\end{equation}
where, F$_1$, and F$_2$ are fluxes observed at time T$_1$ and T$_2$. The flux doubling time or the fastest (or shortest) variability time (t$_{\rm var}$) is the smallest value among available pairs in the total light curve. 
The variability amplitude and the variability time together characterized the variability of the source in various wavebands. The source is observed to be the most variable in X-ray with variability amplitude 1.84$\pm$0.02 and the fastest variability time of the order of half a day (11.28 hours). During the same period, $\gamma$-ray has a variability time of 0.36 days (8.64 hours) with variability amplitude 0.91$\pm$0.13.

\section{Broadband SED modeling}
Simultaneous broadband observations of the flaring period are essential to explore the physical processes responsible for broadband emission through the SED modeling. In the past, BL Lac is modeled many times using the one-zone synchrotron and SSC processes (\citealt{Ghisellini1998}, \citealt{Ravasio2002}).
It is also noticed that, in some cases, during the high flux state one-zone SSC model fails to explain the broadband SED, and alternatively, inverse-Compton scattering of external photons are proposed (\citealt{Sambruna1999}, \citealt{Madejski1999}, \citealt{2000AJ....119..469B}). Subsequently, the broadband SED of BL Lac is conventionally modeled with the one-zone external Compton scenario with the broad-line region (BLR) photons (\citealt{2013ApJ...768...54B}). Furthermore, lepto-hadronic modeling is also done, where high magnetic field and high proton powers are required. A fast flare in BL Lac is observed in very high energy (VHE) $\gamma$-ray by MAGIC and VERITAS during 2015 and 2016 and modeled by IC (\citealt{Magic_2020}) and SSC mechanisms (\citealt{Morris2019}).

\begin{figure*}
    \centering
    \includegraphics[scale=0.4]{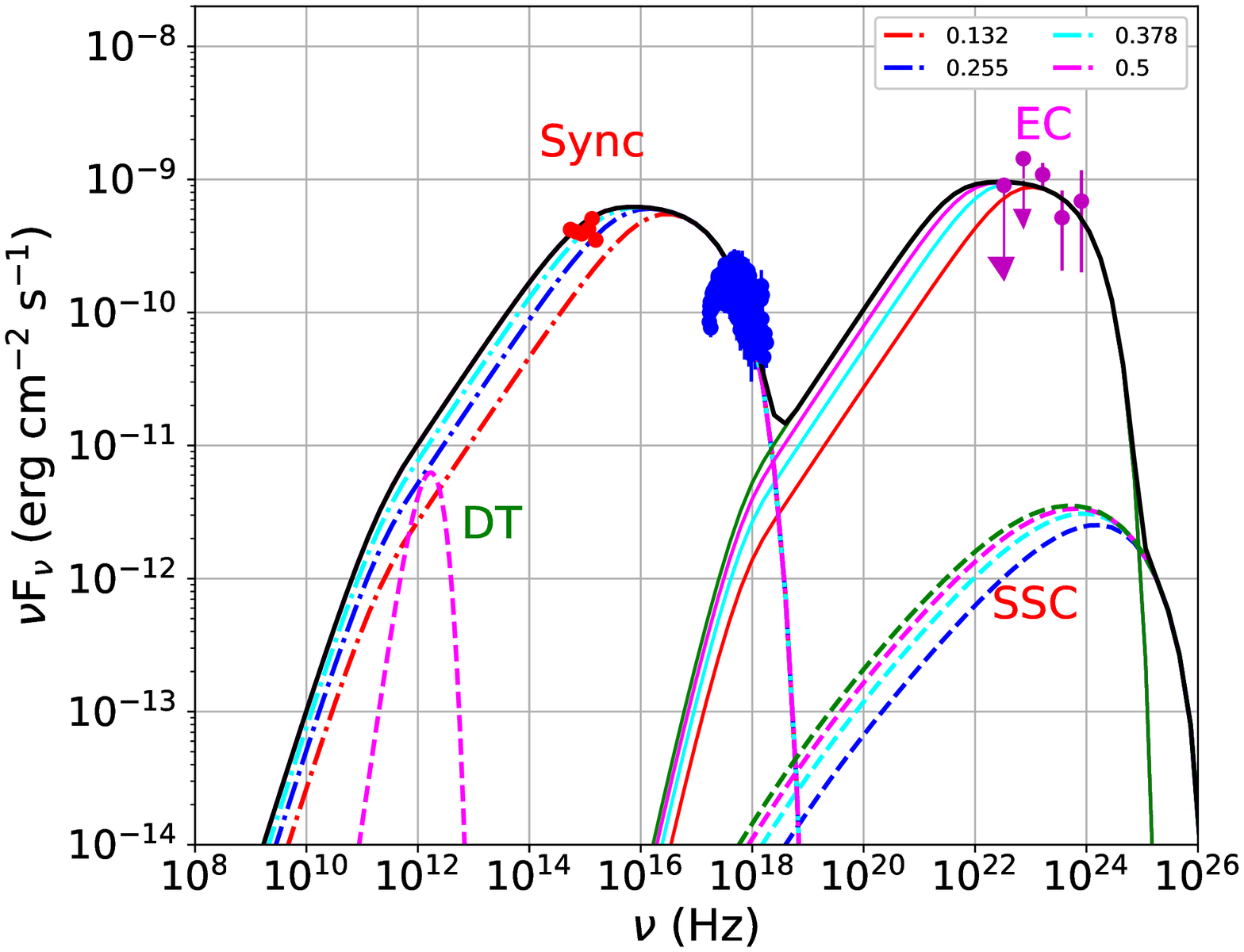}
    \includegraphics[scale=0.4]{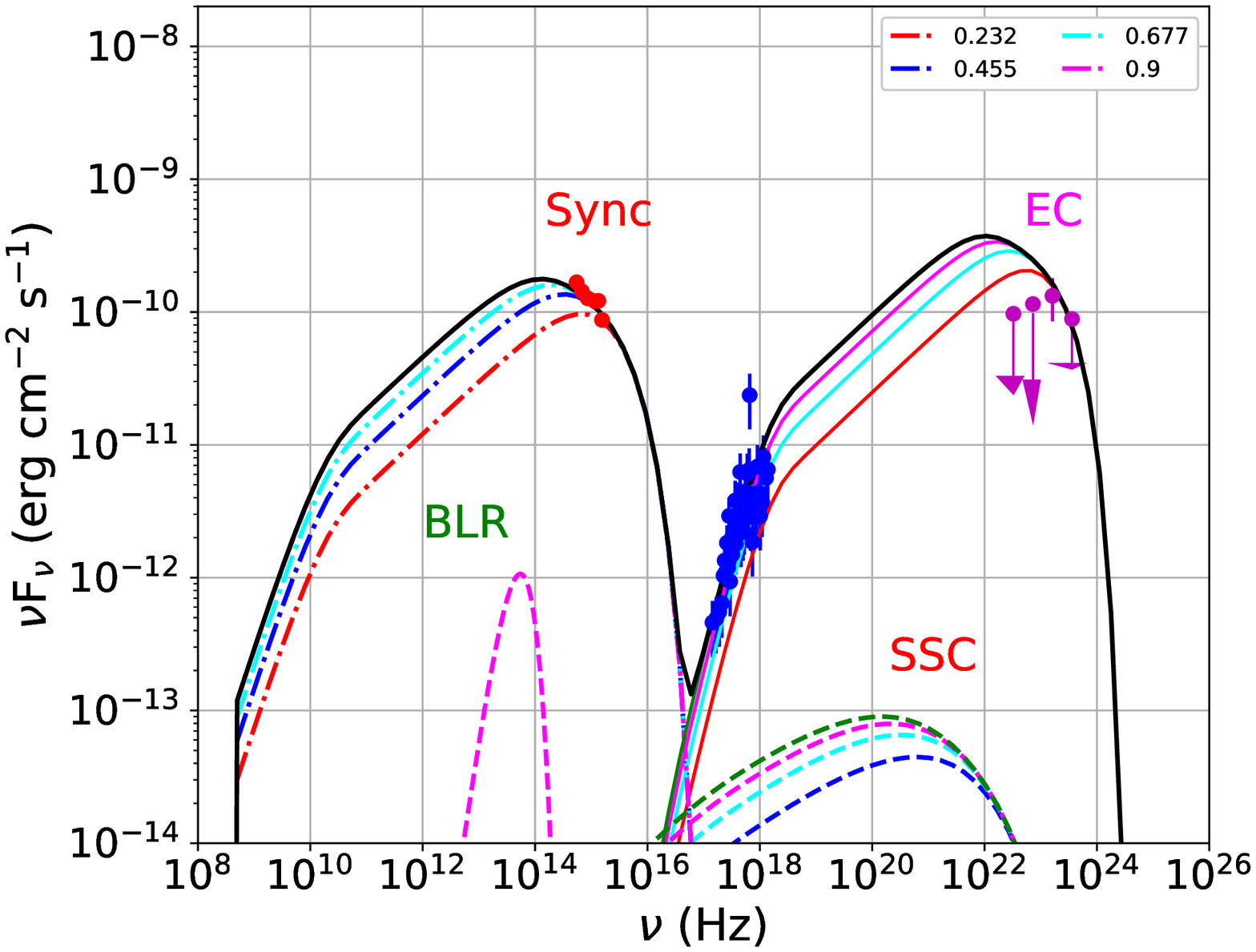}
    \caption{The broadband SED for the two period corresponding to pink dashed line shown in Figure 7. The left panel represents the SED corresponding to the brightest state at MJD 59128.25 and the right panel shows the SED for the low flux state at MJD 59138.17 as shown in Figure 7.  }
    \label{fig:mwsed}
\end{figure*}

\begin{figure}
    \centering
    \includegraphics[scale=0.4]{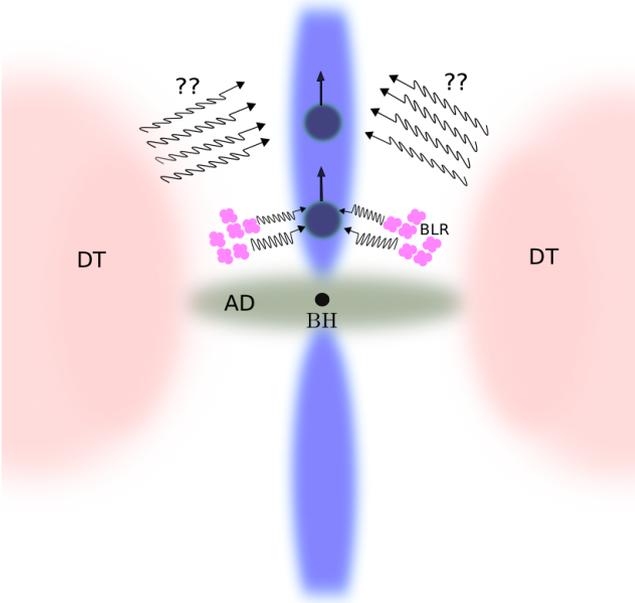}
    \caption{Schematic diagram to show the possible broadband emission scenarios happening during the current 2020 low and high flux state of BL Lac. AD:Accretion disk, BH:Black Hole, DT: Dusty Torus. It is a schematic representation so the image is not to scale.}
    \label{fig:cartoon}
\end{figure}

For the comparison purpose, we modeled the broadband SED of two instances as marked in Figure \ref{fig:MWL} by a vertical pink dashed line. These two states are chosen because of their different X-ray and optical-UV spectral states. 
Optical-UV and X-ray SED are produced for a single observation at MJD 59128.25, and the $\gamma$-ray SED for a duration of half a day centered around the MJD 59128.25.
Similarly, the SED data points for the last observation at MJD 59138.17 (see Figure \ref{fig:MWL}) is produced.
Both the SED is modeled with publicly available code \texttt{GAMERA} (see \citealt{Prince_2020} for details). 
During the high flux state (at MJD 59128.25), the low-energy peak of the broadband SED is constrained by the synchrotron process. To explain the high energy peak, an ambient medium of external photons field is assumed. A similar assumption is also made for blazar OJ 287 by \citep{kush13} to explain the broadband emission during the flaring state (2009 flare). It has been seen that the SSC mechanism often fails to explain the $\gamma$-ray emission during the high flux state of the sources such as OJ 287 and BL Lac. The flare of 2015 in BL lac was modeled under the same assumption by \citet{Magic_2020} where BLR is assumed to be a source of external photons field for the IC scattering. They have assumed a two-zone emission model to explain the broadband SED and their emission regions are located within the BLR.
The broadband SED modeling of the October 2020 flare, in the current work, suggested that the one-zone SSC model is not sufficient to explain the high energy peak of the SED. Therefore, an ambient photon field of energy density 0.62 erg/cm$^3$ and temperature of 310 K (more likely a dusty torus) is assumed.
Similarly, the emission region is chosen close to the BH within the BLR to explain the broadband emission during the low state.
The BLR energy density is estimated by using the L$_{BLR}$ (2.5$\times$10$^{42}$ erg/s) and R$_{BLR}$ (2$\times$10$^{16}$ cm) from the \citet{Magic_2020}. It is found to be $\sim$ 1.6 erg/cm$^{3}$ by following the equation given in \citet{Prince_2020}, Lorentz factor, $\Gamma$=10 (\citealt{Hovatta2009}), and the temperature around 10$^{4}$ K are assumed. Following the relation, R $\sim$ c t$_{var}$ $\delta$/(1+z), the size of the emission region is derived to be $\sim$2.0$\times$10$^{16}$ cm for the $\gamma$-ray variability time, t$_{var}$=0.36 days, and Doppler factor, $\delta$=20 (\citealt{Magic_2020}). 
A log parabola electron distribution is considered to model the broadband SED. The same size of the emission region (2.0$\times$10$^{16}$ cm) is used to explain the low and high state broadband emission but placed at two different locations along the jet axis.
For the high state, the emission region is located far down the jet beyond the BLR, where an ambient photon field of temperature 310 K and energy density 0.62 erg/cm$^3$ is present, probably the case of a dusty torus or the narrow-line region. Though we do not have any observational evidence for that, it is commonly used nowadays to model the broadband SED of FSRQ (\citealt{Sikora2008}). It has also been used for the BL Lac type of source such as OJ 287, \citet{kush13} has suggested a presence of an ambient medium of 250 KeV around the jet and possibly responsible for $\gamma$-ray emission through EC process. Considering the possibility of DT, we estimated its size R$_{DT}$ $\sim$ 3.9$\times$10$^{17}$ cm by using the relation given by \citet{Ghisellini_2009} and disk luminosity, L$_{disk}$ $\sim$2.0$\times$10$^{43}$ erg/s assuming L$_{disk}$ = 10 L$_{BLR}$ from \citet{Maraschi2003}. However, the disk luminosity in \citet{Maraschi2003}
is estimated as $\sim$2$\times$10$^{42}$ erg/s considering the L$_{BLR}$ = $\sim$2$\times$10$^{41}$ from \citet{Celotti1997}. Both estimations of disk luminosity suggest a very weak accretion disk in BL Lac. The size of the possible DT constrained the location of the emission region, suggesting within the DT.
To examine their actual presence, spectral line studies are required for the same period. Currently, no information is available regarding that. Though we have checked the Steward observatory data available for this source for almost 10 years, we did not find any evidence of emission lines in its optical spectrum. However, the SDSS spectra of more than 1000 Fermi blazars are studied in \citet{Paliya_2021} suggesting weak emission lines in blazar BL Lac.
During the high state, the location of the emission region beyond BLR could be justified by the fact that very high-energy photons are observed during this period. As we showed, more than 100 GeV photons are detected during the high state, implying that the emission region is located outside the BLR (BLR act as opaque for $>$20 GeV photons) to avoid the photon$-$photon absorption of high energy $\gamma$-rays. A similar assumption is also made in \citet{Magic_2020} with the bigger emission region located outside the boundary of the BLR. In \citet{Magic_2020}, they considered the scenario with one smaller region located within the BLR under the fact that the BLR is weak in BL Lac (\citealt{Corbett1996, Corbett2000}, \citealt{Capetti2010}) to absorb the high energy $\gamma$-ray, but can provide the external seed photons for the Compton scattering. However, during the low state, no high-energy photons above 10 GeV are observed.
Furthermore, the emission region in the low state is placed close to the BH within the BLR, where the external seed photons from BLR get up-scattered by the high energy electrons and produced the $\gamma$-ray photons. A possible schematic diagram of the physical processes happening during the low and high state is shown in Figure \ref{fig:cartoon}, with distances are not to scale.
A blob is placed within the BLR to explain the low state, and the same size emitting zone is placed outside the BLR to explain the high state of the source. As discussed above, the source of seed photons for the IC scattering in the outer blob are thermal photons of 310 K, which resembles the possible presence of DT in the blazar. However, since it is not confirmed, we mark a question in the schematic representation. 
Under these scenarios, SED is generally well explained (see Figure \ref{fig:mwsed}), although with significant local deviations in the UV and soft X-ray bands.
The corresponding model fit parameters are presented in Table \ref{tab:sed_param}. 
The SED fit parameters such as minimum and maximum energy of electrons, the strength of the magnetic field, the slope of injected electron distributions, and the size of the emission region derived are a bit different than what is reported by \citet{Magic_2020}, and a possible reason could be the strength of the flare during October 2020. 
The difference in the values of the parameters reported in this article and the \citet{Magic_2020} suggest that the source behavior is very complex and difficult to understand. Many more future studies would be required to draw any kind of shared conclusions between BL Lac and any other blazar source. 

\begin{table*}
\centering
\caption{Multi-wavelength SED modeling results with the best fitted parameters values. The input injected electron distribution is LogParabola with reference energy 60 MeV. The Doppler factor and the Lorentz factor are fixed at 20.0 and 15.0 respectively. }
 \begin{tabular}{c c c c c}
 \hline \noalign{\smallskip}
 States & Parameters & Symbols & Values & Period \\
\noalign{\smallskip}  \hline  \noalign{\smallskip}
 OBS 00034748023 (high state) & &&& 0.5 day\\
 & Size of the emitting zone& R & 2.0$\times$10$^{16}$ cm & \\
 & Min Lorentz factor of emitting electrons & $\gamma_{min}$& 20.0 &\\
 & Max Lorentz factor of emitting electrons & $\gamma_{max}$& 4.5$\times$10$^{4}$ &\\
 & Input injected electron spectrum (LP) & $\alpha$ & 1.75 & \\
 & Curvature parameter of the PL spectrum & $\beta$& 0.02 & \\
 & Magnetic field in emitting zone & B & 3.2 G & \\
 & External field energy density & U$_{ext}^{'}$ & 0.62 erg/cm$^3$ & \\
 & External field temperature & T$_{ext}^{'}$ & 310 K & \\
 & Jet power in electrons & P$_{j,e}$ & 7.57$\times$10$^{43}$ erg/s & \\
 & Jet power in magnetic field & P$_{j,B}$ & 3.46$\times$10$^{45}$ erg/s & \\
 & Jet power in protons & P$_{j,P}$ & 2.41$\times$10$^{43}$ erg/s& \\
 & Total jet power & P$_{jet}$ & 3.56$\times$10$^{45}$ erg/s& \\
 \noalign{\smallskip} \hline   \noalign{\smallskip}
OBS 00034748034 (low state) & &&& 1 day \\
& Size of the emitting zone& R & 2.0$\times$10$^{16}$ cm & \\
 & Min Lorentz factor of emitting electrons & $\gamma_{min}$& 5.0 &\\
 & Max Lorentz factor of emitting electrons & $\gamma_{max}$& 3.8$\times$10$^{3}$ &\\
 & Input injected electron spectrum (LP) & $\alpha$ & 2.20 & \\
 & Curvature parameter of the PL spectrum & $\beta$& 0.002 & \\
 & Magnetic field in emitting zone & B & 4.5 G & \\
 & External field energy density & U$_{BLR}^{'}$ & 1.6 erg/cm$^3$ & \\
 & External field temperature & T$_{BLR}^{'}$ & 1$\times$10$^4$ K & \\
 & Jet power in electrons & P$_{j,e}$ & 5.80$\times$10$^{43}$ erg/s & \\
 & Jet power in magnetic field & P$_{j,B}$ & 6.83$\times$10$^{45}$ erg/s & \\
 & Jet power in protons & P$_{j,P}$ & 2.34$\times$10$^{44}$ erg/s& \\
 & Total jet power & P$_{jet}$ & 7.13$\times$10$^{45}$ erg/s& \\
\noalign{\smallskip}  \hline  \noalign{\smallskip}
 \end{tabular}
 \label{tab:sed_param}

\end{table*}

\subsection{Jet Power}
The total power carried by jet and by the individual components (leptons, protons, and magnetic field) can also be estimated by using the relation,

\begin{equation}
	 P_{\text{jet}}=\pi R^2 \Gamma^2 c	(U^\prime_e+U^\prime_B+U^\prime_p)
	 \label{equ:Total_Power}
\end{equation}
where $\Gamma$ is the bulk Lorentz factor; $U^\prime_B$, $U^\prime_\text{e}$, and $U^\prime_\text{p}$ are the energy density of the magnetic field, electrons (and positrons), and cold protons respectively in the co-moving jet frame (primed quantities are in the co-moving jet frame while unprimed quantities are in the observer frame). The power in leptons is given by

\begin{equation}
	P_{\text{e}}= \frac{3\Gamma^2 \, c}{4R} \int_{E_{\text{min}}}^{E_{\text{max}}} E \, Q(E) \, dE 
\end{equation}
where, $Q(E)$ is the injected particle spectrum. The integration limits, $E_{\text{min}}$ and $E_{\text{max}}$ are calculated by multiplying the minimum and maximum Lorentz factor ($\gamma_{min}$ and $\gamma_{max}$) of the electrons with the rest-mass energy of electron respectively.

The power in magnetic field is calculated using
\begin{equation}
    P_B= R^2 \Gamma^2 c \frac{B^2}{8}
\end{equation}
where $B$ is the magnetic field strength obtained from the SED modeling. To calculate $U^\prime_p$ , we assumed the ratio of the electron-positron pair to the proton is 10:1 with maintaining the charge neutrality condition. The total jet power in protons is calculated by calculating the total energy led by protons and the volume of the emission region. The power estimated in all components is shown in Table \ref{tab:sed_param} and the total jet power in both the cases are the order of 10$^{45}$ erg/s. We observed that the jet has more power in the magnetic field than the injected electron, suggesting the jet is dominated by magnetic power. A significant spectral change in the injected electron spectrum is observed between low and high states, suggesting the involvement of two different populations of electrons. More electron energy during high state and different spectral index of electrons population suggest that the flare is produced with a newly fresh electron, and as a result, more electron power in flaring state compared to low state. 
A big sample of FSRQ and BL Lac is studied by \citet{Ghisellini_2015}, suggesting the ratio L$_{disk}$/L$_{Edd}$ as $\sim$0.1 and 0.01 for them, respectively.
Using the ratio of BL Lac type sources, the Eddington power is estimated as an order of $\sim$10$^{45}$ erg/s, which is consistent with the total power estimated during the high and low state.
However, considering the mass of the SMBH of BL Lac, M$_{\rm SMBH}$=1.58$\times$10$^{8}$M$_{\odot}$ from \citet{Wu2018}, we estimated the Eddington luminosity of the source, L$_{Edd}$=1.3$\times$10$^{38}$(M$_{\rm SMBH}$/M$_{\odot}$) such as , L$_{Edd}$ =2.05$\times$10$^{46}$ erg/s, an order of magnitude larger than the total jet power estimated from the SED modeling.

\section{Change in location of synchrotron peak: HBL like component}
This section is dedicated to examining any change in the location of the synchrotron peak, constrained by the NIR to optical-UV and X-ray emissions.
NIR data is not available for the current flaring state, and hence optical-UV and X-ray spectral points are considered to identify any shift in the synchrotron peak.
In the past, the log-parabolic model is used to fit the NIR to X-ray SED data points in order to estimate the exact location of the synchrotron peak (\citealt{Massaro2004}, \citealt{Kapanadze_2018}),
in addition, \citet{Raiteri2013} used the log-cubic model. BL Lac is a TeV source, and the synchrotron peak of this source along with many other TeV blazars was estimated by \citet{Nilsson2018} by fitting the total SED with two log-parabolic spectra, where it is classified as LBL type blazar.
The SED constructed in \citet{Raiteri2009} with the simultaneous observation from radio to X-ray suggests that the synchrotron peak resides in the near-IR band. They also compared their SED with the previous observations from other X-ray instruments and concluded that the X-ray spectrum is highly variable, and it changes from extremely steeper to extremely harder and more or less curved in the intermediate state.
Similar results are found in this study with Swift-XRT at two different states. In the low state, the spectrum is very hard, followed by an extremely steep nature in the high state with changing spectral shape. 
 Along with the X-ray, the optical-UV spectrum also changed its shape, and combined together; the synchrotron peak appears to be shifted towards the higher energy during the high state. The peak of the synchrotron emission is shown in Figure \ref{fig:mwsed}, during the high state, it peaks at $\sim$10$^{16}$ Hz, and that is the case for HBL type sources (\citealt{Padovani1995}, \citealt{Abdo2010}). On the other hand, during a low flux state, the synchrotron emission peaks around $\sim$10$^{14}$ Hz, suggesting source an IBL type blazar. These findings clearly suggest that the source changed its spectral characteristic from IBL to HBL when transiting from low to high flaring in October 2020. Similar results are also observed for other TeV blazar, Mkn 501, by \citet{Anderhub2009}.

\section{Summary and Conclusions}
In this article, we reported the brightest X-ray and optical flares ever observed from blazar BL Lac along with the broadband light curves.
Before the bright X-ray activity, the source was reported to be flaring in $\gamma$-ray and was also detected in very high energy $\gamma$-ray by MAGIC as reported in many Atels (\citealt{ATel14032,Blanch2020}).
Our study suggests the significant spectral change in optical-UV as well as in the X-ray spectrum. During the low flux state of the source, the spectrum was observed to be harder; however, at the high flux state, extreme softness in the X-ray spectrum is seen. For the first time, a "softer-when-brighter" trend is seen, while the usual "harder-when-brighter" trend is reported previously by \citealt{Raiteri2009}. 
We extended the $\gamma$-ray data collection beyond the current X-ray flaring state, and it is again observed to be flaring in $\gamma$-ray while X-ray is in a low flux state. Contemporaneous $\gamma$-ray flare with respect to X-ray is defined as \texttt{state 1} and a "softer-when-brighter" trend is seen in $\gamma$-ray. On the other hand, \texttt{state 2} of the $\gamma$-ray flare showed a reverse trend ("harder-when-brighter"). Changes in the $\gamma$-ray spectral behavior during two different occasions suggest that these two flares are produced under different circumstances and probably associated with different processes. During the $\gamma$-ray \texttt{state 1} or the X-ray flaring state, a very high energy photon of energy order of $\sim$100 GeV are observed, suggesting emission is produced outside the BLR to avoid the photon-photon absorption.
The source showed the highest flux variability in X-ray and $\gamma$-ray followed by the UV and optical with the fastest variability time in $\gamma$-ray and X-ray of the order of half a day. We also examined the change in X-ray spectral shape during the high and low states. 
A harder spectrum in the low state is followed by a flat spectrum in an intermediate state, and eventually, a very steep spectrum in the high state is observed.
Similar trends are seen in the optical-UV spectrum also (Figure \ref{fig:xray-uv}). The simultaneous broadband light curves showed that the flares in various bands are not peaking at the same time, rather lagging with each other. The DCF study suggests optical-UV emission leads the X-ray emission and X-ray emission leads the $\gamma$-ray emission by order of the day.
We modeled the broadband SED of two states, e.g., low state and high state, with a one-zone emission model. For the low state, the emission region is placed within the BLR, and for the high state is located outside the BLR to avoid the photon-photon absorption. 
During the high state, the synchrotron peak is observed at $\sim$10$^{16}$ Hz, suggesting an HBL type behavior.
We concluded that when the source moves from the low flux state to the high state, the spectral behavior changes from IBL to the HBL type. Such behavior is seen for the first time in blazar, BL Lac. To confirm its oscillatory nature between IBL and HBL, many more future studies would be required.  



\section*{Acknowledgements}

R.P. thank Swayamtrupta Panda for scientifc discussions and helping with Cartoon shown in paper. R.P. also thank Prof. Bozena Czerny for scientific discussion. R.P. thank Rukaiya Khatoon and Gayathri Raman for the proof reading and discussions. This work has made use of publicly available Fermi data obtained from FSSC and Swift data from Neil Gehrels observatory. This research has also made use of XRT data analysis software (XRTDAS) developed by ASI science data center, Italy. 
      The project was partially supported by the Polish Funding Agency National Science Centre, project 2017/26/A/ST9/00756 (MAESTRO 9), and MNiSW grant DIR/WK/2018/12.

\section*{Data Availability}

This research has made use of archival data from various sources e.g.Fermi and Swift observatory and their proper links are given in the manuscript. All the models and software used in this manuscript are also publicly available.

\bibliographystyle{mnras}
\bibliography{reference-list.bib}








\bsp	
\label{lastpage}
\end{document}